\documentclass[12pt]{iopart}
\usepackage{braket}
\usepackage{graphicx}
\usepackage{hyperref}

\begin{document}

\title[Nonlinearity accelerates thermalization of the quartic FPUT with stochastic baths]{Nonlinearity accelerates the thermalization of the quartic FPUT model with stochastic baths}

\author{Harald Schmid}
\address{Universit\"at Regensburg, GER-93040 Regensburg, Germany}
\address{Scuola Normale Superiore, I‐56126 Pisa, Italy}
\ead{haraldschmid777@gmail.com}
\author{Sauro Succi}
\address
{
Center for Life Nano Science @ La Sapienza, Italian Institute of Technology, I-00161 Roma, Italy
}
\address{Scuola Normale Superiore, I‐56126 Pisa, Italy}
\author{Stefano Ruffo}
\address{
SISSA, Via Bonomea 265, I-34136 Trieste, Italy
}
\address
{
INFN Trieste, I-34149 Trieste, Italy
}

\vspace{10pt}

\begin{abstract}
We investigate the equilibration process of the strongly coupled quartic Fermi-Pasta-Ulam-Tsingou (FPUT) 
model by adding Langevin baths to the ends of the chain. 
The time evolution of the system is investigated by means of extensive
numerical simulations and shown to match the results expected from equilibrium 
statistical mechanics in the time-asymptotic limit.
Upon increasing the nonlinear coupling, the thermalization of the energy spectrum 
displays an increasing asymmetry in favour of small-scale, high-frequency modes, which 
relax significantly faster than the large-scale, low-frequency ones. 
The global equilibration time is found to scale linearly with system size and shown 
to exhibit a power-law decay with the strength of the nonlinearity and temperature. 
Nonlinear interaction adds to energy distribution among modes, thus speeding 
up the thermalization process.
\end{abstract}

\vspace{2pc}

\section{Introduction}

Thermalization in the Fermi-Pasta-Ulam-Tsingou (FPUT) model has attracted much attention
since the original formulation of the problem \cite{Fermi1955,Gallavotti2007,Berman2005}. 
In the weak coupling regime the system reaches equipartition on a timescale which depends 
as a power-law \cite{Benettin2013,Onorato2015,Lvov2018} on the energy density.
In this work, we take a different approach by studying the equilibration process ranging 
from the weak to the strong coupling regime. In particular, we explore the steady-state dynamics 
and the relaxation to steady-state of the quartic FPUT attaching stochastic Langevin 
baths to both ends of the chain. 
A detailed study of the nonlinear model in equilibrium conditions was performed 
in \cite{Livi1987} by realizing the canonical setting in a textbook manner, considering
 a small part of the microcanonical system.
We show that our description
using Langevin baths proves to be a successful framework to simulate 
both equilibrium and non-equilibrium properties of
the nonlinear model in a
canonical setting.
We find by numerical simulations of the stochastic equations that 
increasing the nonlinear coupling accelerates the approach to 
equilibrium. Even at high nonlinearity equilibrium is surprisingly 
characterized by quasi-equipartition of linear energy among Fourier modes 
and relaxation to equilibrium is faster for high-frequency modes. We 
derive analytically a compact representation of the equilibrium 
properties of the model, introducing a dimensionless scaling variable
$\lambda k_B T$ with temperature of the baths $T$ and coupling
strength $\lambda$. The expressions obtained from the canonical partition function 
for equilibrium quantities, such as internal energy and nonlinear energy, are 
in satisfactory agreement with the numerical results of the stochastic model.
The equilibration time is shown to be linear in the system-size for a sufficiently 
large number of oscillators and is probed as a function of coupling (temperature) 
at different temperatures (couplings). We find that both dependencies are well-fitted by
a decaying power-law with exponent $-1/3$ across four decades in temperature. 

\section{The model}

We consider the following extension to the quartic FPUT with $N$ particles 
given by the Langevin equations  \cite{Lepri2003}

\begin{equation}
\label{eq:beta-LFPU}
\eqalign{
\ddot{q}_j=& q_{j+1}+q_{j-1}-2q_j+\lambda\left[\left(q_{j+1}-q_{j}\right)^3-\left(q_{j}-q_{j-1}\right)^3\right]
\cr
&-\delta_{1,j}\left[\gamma \dot{q}_j-\eta_L(t)\right]-\delta_{N,j}\left[\gamma \dot{q}_j-\eta_R(t)\right],
}
\end{equation}

\noindent where the baths at the left and right edge ($L,R$) have the same
temperature $T$

\begin{equation}
\fl \braket{\eta_a(t)}=0,
\hspace{0.5cm}
\braket{\eta_a(t)\,\eta_b(t')}=2\gamma\,k_B T\,\delta_{a,b}\,\delta(t-t'), 
 \hspace{0.5cm} a,b = \{L,R\}.
\end{equation}

\begin{figure}[t!]
\includegraphics[width=1\textwidth]{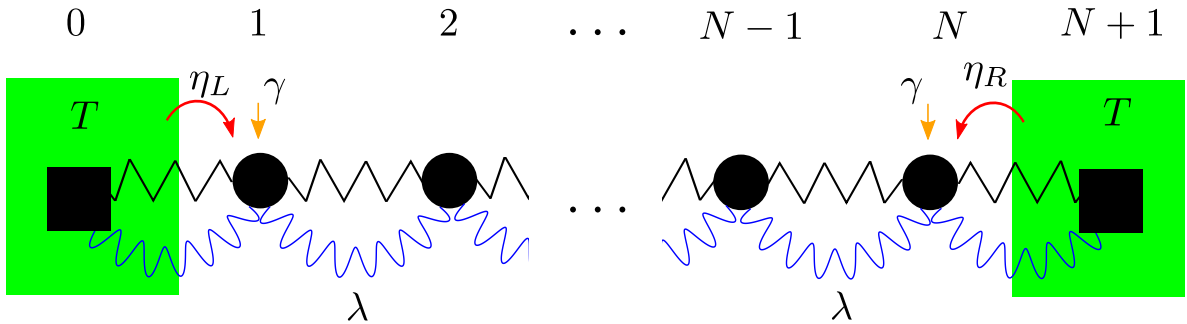}
\caption{Sketch of the model (\ref{eq:beta-LFPU}). A chain of $N$ particles with quadratic (black springs) and quartic interaction (blue whirls, coupling strength $\lambda$) is attached to stochastic baths of the same temperature $T$. The ends ($0$ and $N+1$) are fixed.}
\label{fig:model_sketch}
\end{figure}

\noindent Fixed boundary conditions were imposed such that $q_0=q_{N+1}=0$. Initially, the position and momenta are plain for all sites, $q_j(0)=0$ and $\dot{q}_j(0)=p_j(0)=0$.
For reasons of simplicity, the mass $m$ and the spring constant of the harmonic interaction $k$  are set to unity, i.e.~$m=1$ and $k=1$. Having a physical realization of our system in mind \cite{Goossens2019,Pierangeli2019}, we couple stochastic motion only to the edges of the chain. The Langevin approach for the baths is a special case of the general Markovian evolution of the baths where the reservoirs are not affected by the system at all. The particular choice of the baths provides a physical implementation of the  thermostats compared with more efficient schemes like Nosé-Hoover baths \cite{Nose1984,Hoover1985} which have been applied to determine probability distributions for canonical momenta \cite{Demirel1997}. The model without nonlinearity ($\lambda=0$) coupled to Langevin baths is known to thermalize, i.~e.~all particles have equal kinetic energy as time goes to infinity \cite{Rieder1967,Kim2017}.

\section{Numerical implementation}

We have integrated the equations of motion using a standard fourth order Runge-Kutta scheme (RK4) 
for the deterministic part and a simple first order Euler–Maruyama method for 
the stochastic term \cite{numerical_recipes}.  
Other integration schemes (e.g. Velocity-Verlet), have also been considered for the deterministic 
part, which proved to be slower at a comparable order of accuracy. 
Setting the dissipation to unity ($\gamma=1$ fixed for the rest of the paper), a 
stepsize $\Delta t=0.01$ proved to be stable and sufficiently accurate for the purpose of this paper. 
Considering the deterministic part separately ($T=0$), the largest observed deviation from 
energy conservation using the RK4 scheme, was of the order of $10^{-9}$, for the 
case of $N=32$ sites and strong nonlinearity $\lambda=10$. 
For the full stochastic system, equipartition was reached numerically and the 
fluctuation-dissipation theorem was satisfied within less then   1 \% for the 
longest simulation times (5000 timesteps).
Ensemble-averaging and time-averaging were implemented simultaneously for every quantity
under inspection.
In equations:

\begin{equation}
F(t) = \frac{1}{2\,\delta t}\sum\limits_{l=1}^{N_{run}}\int\limits^{t+\delta t}_{t-\delta t} \mathrm{d}s\, f_l(s),
\end{equation}
\noindent with $f_l(t)$ being a measured quantity for a single Wiener-process realization. 
Unless explicitly mentioned otherwise, we will refer from now on to every quantity 
$F$ in the text as the averaged one. 
A window of $\delta t=100$ and $N_{run}=100$ runs are used for the statistics. 
Numerical errors appear due the choice of the timestep which affects predominantly the 
integration of the stochastic part, due to the low-order Euler-Mayurama scheme 
(time-integration is numerically more costly than ensemble-averaging since the latter
can be run in parallel across independent tasks). 
We have checked the validity of our results by doubling the timestep, size of 
the ensemble and of the time-averaging window for the time-average, thus 
confirming satisfactory convergence of the stochastic simulations.
Higher order schemes for the stochastic part might improve the efficiency 
of the code and will be considered in the future for longer time simulations.  

\begin{figure}[b!]
\includegraphics[width=1\textwidth]{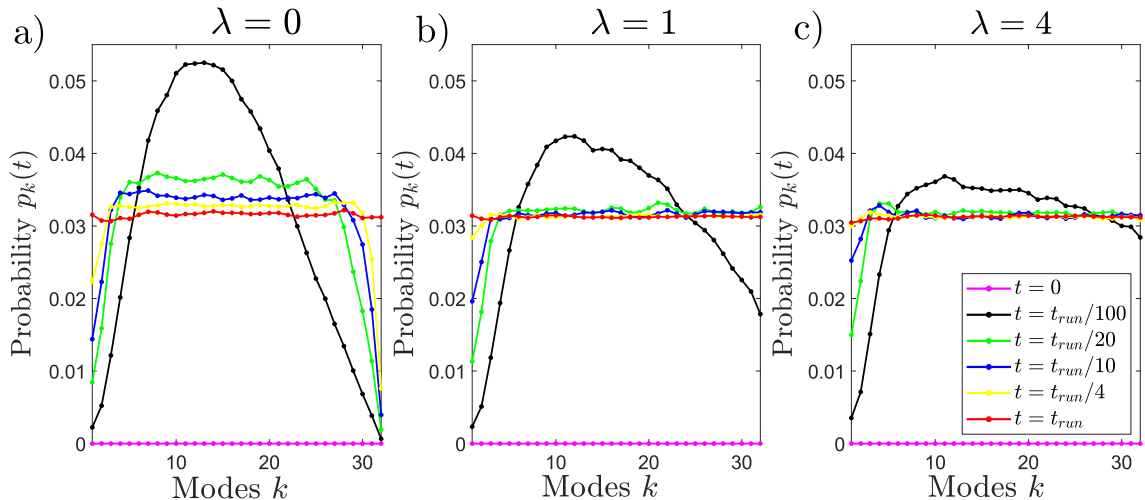}
\caption{Distribution of modes (bath temperature $T=1$, dissipation $\gamma=1$). We evolved the system of $N=32$ sites for $t_{run}=5000$ and three different strengths of the nonlinearity $\lambda$ until the distribution becomes flat due to equipartition. For the linear system ($\lambda=0$) modes around $k\approx 12$ are excited early (black curve). The slowest and fastest modes reach equilibrium latest. Ramping up the coupling  ($\lambda=1,4$) creates an asymmetry in the thermalization between modes at the ends of the spectrum as high-oscillating modes reach equilbrium faster than low-oscillating modes. Furthermore, equipartition is reached faster for larger coupling due to the quicker distribution of energy into modes at both ends of the spectrum.}
\label{fig:distribution_of_modes}
\end{figure}

\begin{figure}[t!]
  \includegraphics[width=1\textwidth]{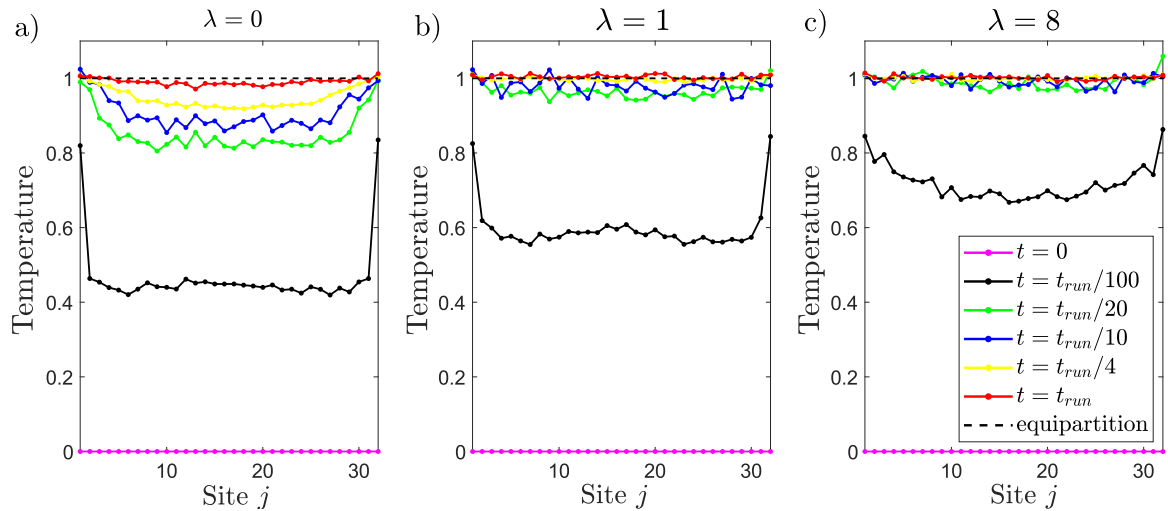}
\caption{Temperature profiles $T_j(t)$ (same parameters as in figure \ref{fig:distribution_of_modes}).  Equipartion is reached faster for larger coupling.}
\label{fig:temperature profiles}
\end{figure}

\section{Time-evolution of the normal modes}

\noindent The kinetic energy of a single site defines a temperature distribution in real space $k_B T_j(t) = \dot{q}_j^2(t)$. For the linear Langevin chain (recovered by setting $\lambda=0$) all oscillators equilibrate to the temperature of the baths, i.~e.~$T_j(\infty)=T$ \cite{Rieder1967,Nakazawa1970}. We will show that this holds also for the nonlinear case. It is useful to introduce the normal modes of the harmonic system
\cite{Ford1992}

\begin{equation}
\label{eq: Normal modes}
A_k(t) = \sqrt{\frac{2}{N+1}}\sum\limits_{j=1}^N q_j(t) \sin\left(\frac{k j\pi}{2\left(N+1\right)}\right).
\end{equation}

\noindent Their frequencies are $\omega_k = 2\sin(k\pi/(2N+2))$ and the linear energy is given by $E_k = 1/2(\dot{A}_k^2+\omega_k^2 A_k^2)$. The total energy stored in the linear motion of the modes is $\sum_k E_k$. The probability to find the system in mode $k$ is adopted according to \cite{Onorato2015,Livi1985,Pistone2019} 

\begin{equation}
\label{eq: probability}
p_k(t) = \frac{E_k}{\sum_k E_k}(t).
\end{equation}

\noindent If the nonlinearity is weak the total energy $E$ is well-approximated by the linear energy in the modes, i.~e.~$E\approx\sum_k E_k$. This is not true anymore for the strong coupling regime. Nevertheless, (\ref{eq: probability}) defines the probability distribution of the normal modes at every value of the coupling, and is properly normalized by $\sum_k E_k$. Excluding an anharmonic term from the definition gives us a clear perspective on the modification of the spectrum when the coupling is changed. We observe that the nonlinearity has an impact on the time evolution of the spectrum $p_k$. 
\\
Snapshots of temperature profiles $T_j(t)$ and the distribution of modes $p_k(t)$ for different coupling strengths are shown at different points during their evolution in figure \ref{fig:distribution_of_modes} and \ref{fig:temperature profiles}. For the latest time in our simulation, the system approaches a homogeneous temperature in real space and a flat spectrum in mode space, regardless of the coupling strength. Comparing distributions/temperature profiles for different coupling strengths at the early points in the evolution, the nonlinearity accelerates the thermalization process. For the linear case ($\lambda=0$), modes around $k\approx N/3$ are fastest excited at early stages in the evolution. In comparison with the real space evolution, the exterior sites $j=1$ and $j=N$ attached to the baths are early excited. Low- and high-oscillating Fourier modes at the ends of the spectrum reach equipartition latest and nearly at the same time. Ramping up $\lambda$, an asymmetry  in the spectrum is observed, as high-oscillating modes are quicker to reach equipartition than the low ones. At $\lambda=4$ (figure \ref{fig:distribution_of_modes} c)), the first snapshot in the evolution (black curve) displays a spectrum which is almost flat for modes $k>8$ and steeply declines for low-oscillating modes. The high-oscillating modes approach equilibrium on the same time scale as the fastest relaxed modes whereas low modes trail behind.

\section{Thermodynamics from the stochastic model }

Next, we derive analytic expressions for the internal energy, and the nonlinear and harmonic part of the energy in equilibrium. They are required to validate the corresponding time-asymptotic quantities in the simulation of the stochastic quartic FPUT. The partition function for the system reads

\begin{equation}
\label{eq:partition function}
\fl
Z =   \int\,\prod\limits_{j=1}^{N}\mathrm{d}p_j\,\mathrm{d}q_j\, \exp\left\{-\left(k_B T\right)^{-1}\left[\sum\limits_{j=0}^{N}\frac{1}{2}\,p_j^2+\frac{1}{2}\left(q_{j+1}-q_{j}\right)^2+\frac{\lambda}{4}\left(q_{j+1}-q_{j}\right)^4\right]  \right\}
\end{equation} 

\noindent with momenta $p_j$ and $p_0=p_{N+1}=0$ as well as $q_0=q_{N+1}=0$. Singling out the kinetic term we make the coordinate transformation \cite{Livi1987}

\begin{equation}
\label{eq:transform}
\fl \phi_L = q_1-q_0,\hspace{0.3cm}
\phi_1 = q_2-q_1,\hspace{0.3cm} \dots\hspace{0.3cm} ,\,\phi_{N-1} = q_N-q_{N-1},\hspace{0.3cm} \phi_{N} = q_{N+1}-q_{N},\hspace{0.3cm} \phi_{R} = q_{N+1}.
\end{equation}

\noindent The variables $\phi_{L}$ and $\phi_L$ are not real coordinates but parameters due to the fixed boundaries. The Jacobian of the transformation is invertible and has determinant equal to the identity. The partition function is transformed to 

\begin{equation}
Z = \xi(N) \,\left(\pi k_B T\right)^{N/2}\int\prod\limits_{j=1}^{N}\mathrm{d}\phi_j\, \exp\left\{-\left(k_B T\right)^{-1}\sum\limits_{j=0}^{N}\left[\frac{1}{2}\,\phi_j^2+\frac{\lambda}{4}\, \phi_j^4 \right] \right\},
\end{equation}

\noindent displaying the same contribution of every variable $\phi_j$. The prefactor $\xi(N)=\exp\left(-\frac{\left(q_0-q_1\right)^2}{N+1}\right)/\sqrt{N+1}$ is a relict of the transformation and does not depend on the coordinates due to the fixed boundary conditions. The integral of the quartic exponential is found in terms of modified Bessel functions of the second kind $K_\nu(z)$. The partition function becomes

\begin{equation}
Z = \xi(N)\left[\frac{\pi k_B T}{2\lambda}\,\exp\left(\frac{1}{8 k_B T \lambda }\right)K_{\frac{1}{4}}\left(\frac{1}{8 k_B T \lambda }\right)\right]^{N/2}.
\end{equation}

\begin{figure}[t!]
\centering
\includegraphics[width=1\textwidth]{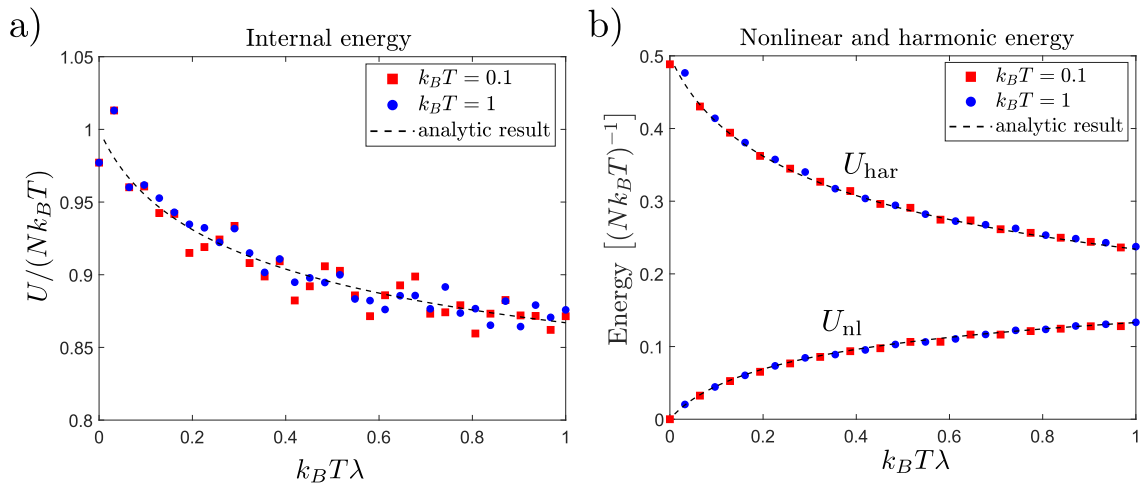}
\caption{Dependence of the energy of the quartic FPUT in equilibrium ($N=32$ sites). a) The internal energy normalized by the thermal energy of a linear chain, i.e. $U/(Nk_B T)$, is a function only of the product $k_B T \lambda$. The simulation using the stochastic baths agrees with the analytic result (\ref{eq:internal energy beta-LFPU}). b) Nonlinear energy $U_{\mathrm{nl}}$ increases with coupling while the harmonic energy $U_{\mathrm{har}}$ decreases (\ref{eq:harmonic and nonlinear energy}) (black dotted).}
\label{fig:energy}
\end{figure}

\noindent This is equivalent to the result given in \cite{Livi1987} by the relation of the parabolic cylinder functions $D_\nu(y)$ and the Bessel functions $D_{-1/2}(y)= \sqrt{y/(2\pi)}\,K_{1/4}\left(y^2/4\right)$, upon changing from fixed to periodic boundary conditions. 
\\
 The internal energy is related to the partition function by $U=k_B T^2\,\partial/\partial T\,\log(Z)$. We obtain

\begin{equation}
\label{eq:internal energy beta-LFPU}
U = \frac{Nk_B T}{4}+\frac{N}{8\lambda}\mathcal{K}\left(\frac{1}{8k_BT \lambda}\right)
,
\end{equation}

\noindent where we have definded the ratio of the Bessel functions

\begin{equation}
\mathcal{K}(y)=\frac{
K_{\frac{5}{4}}\left(y\right) 
} 
{K_{\frac{1}{4}}\left(y\right) 
}-1.
\end{equation}

\noindent Again, the result is equivalent to \cite{Livi1987}. Dividing in (\ref{eq:internal energy beta-LFPU}) by $k_BT$, the internal energy has a dimensionless scale $z=8k_BT\lambda$ 

\begin{equation}
\frac{U}{Nk_BT}(z) = \frac{1}{4}+\frac{1}{z}\,\mathcal{K}\left(\frac{1}{z}\right).
\end{equation}

\noindent The coupling $\lambda$ plays by the found relation for $z$ the role of an inverse nonlinear temperature scale. It is also worth investigating the harmonic part and  nonlinear part of the energy 

\begin{eqnarray}
U_{\mathrm{har}} =&\,Z^{-1}
  \int\,\prod\limits_{j=1}^{N}\mathrm{d}p_j\,\mathrm{d}q_j\,\sum\limits_{j=0}^{N}\frac{1}{2}\left(q_{j+1}-q_{j}\right)^2 \,\exp\left\{-\left(k_B T\right)^{-1} H\right\},
  \\ 
 U_{\mathrm{nl}}  =&\,Z^{-1}
  \int\,\prod\limits_{j=1}^{N}\mathrm{d}p_j\,\mathrm{d}q_j\,\sum\limits_{j=0}^N\frac{\lambda}{4}\left(q_{j+1}-q_{j}\right)^4 \,\exp\left\{-\left(k_B T\right)^{-1} H\right\},
\end{eqnarray}

\noindent where $H$ is the Hamiltonian of the quartic FPUT. The change of variables (\ref{eq:transform}) and rewriting by parameter differentiation yields

\begin{eqnarray}
U_{\mathrm{har}} =& -\lambda k_B T \frac{\partial}{\partial\lambda}
\log\left[\int\mathrm{d}\phi\,\exp\left\{-\left(k_B T\right)^{-1}\left(\frac{1}{2}\,\phi^2 +\frac{\lambda}{4}\, \phi^4 \right) \right\}\right]^{N},
\\
 U_{\mathrm{nl}} = &-\lambda k_B T \frac{\partial}{\partial g}\bigg|_{g=1}
\log\left[\int\mathrm{d}\phi\,\exp\left\{-\left(k_B T\right)^{-1}\left(\frac{g}{2}\,\phi^2 +\frac{\lambda}{4}\, \phi^4 \right) \right\}\right]^{N}.
\end{eqnarray}

\noindent The result can again be given in dimensionless form

\begin{equation}
\label{eq:harmonic and nonlinear energy}
 \frac{U_{\mathrm{har}} }{Nk_B T}(z)  =\frac{2}{z}\,\mathcal{K}\left(\frac{1}{z}\right)-1,\,\quad
\frac{ U_{\mathrm{nl}} }{Nk_B T}(z) = \frac{3}{4}-\frac{1}{z}\, \mathcal{K}\left(\frac{1}{z}\right).
\end{equation}

\noindent Observe that $z^{-1}\mathcal{K}\left(z^{-1}\right)\rightarrow \frac{1}{2}$ for infinitely strong coupling $z\rightarrow \infty$, so the asymptotics is given by

\begin{equation} 
\label{eq:asymptotics}
\fl \lim\limits_{k_B T \lambda\rightarrow \infty} U = \frac{3}{4}N\,k_B T, \hspace{1cm}
\lim\limits_{k_B T \lambda\rightarrow \infty}  U_{\mathrm{nl}}  = \frac{N}{4}\,k_B T,\hspace{1cm}
\lim\limits_{k_B T \lambda\rightarrow \infty}U_{\mathrm{har}} = 0.
\end{equation}

\noindent The asymptotic (and maximum) ratio of nonlinear energy $U_{\mathrm{nl}} $ and internal energy $U$ settles at  $ U_{\mathrm{nl}} /U=\frac{1}{3}$. The following ratio yields the contribution of nonlinear coupling to potential energy 

\begin{equation}
\label{eq:nonlinear coupling}
\eta(z) = \frac{ U_{\mathrm{nl}}  }{U_{\mathrm{har}} + U_{\mathrm{nl}}   } = \frac{3z-4\,\mathcal{K}\left(z^{-1}\right)}{4\,\mathcal{K}\left(z^{-1}\right)-z}.
\end{equation} 

\noindent It assumes values on the unit interval, i.e. $\eta(0)=0$ (purely linear) and $\eta(\infty)=1$ (purely nonlinear), and thus provides a good measure to separate strong from weak coupling which we define by

\begin{equation}
\eta(z) \ll 1, \hspace{0.5cm} \mathrm{weak\, coupling}
.
\end{equation}

\noindent We have simulated the linear case $\eta=0$ up to values $\eta\approx 0.36$ (at $k_B T \lambda =1$), covering both weak and strong coupling regime. The stochastic implementation of the canonical ensemble of the quartic FPUT shows quantitative agreement with the values of internal, linear and nonlinear energy expected from equilibrium statistical mechanics (see figure \ref{fig:energy}).

\section{Equilibration time}

\begin{figure}[b!]
\centering
\includegraphics[width=1\textwidth]{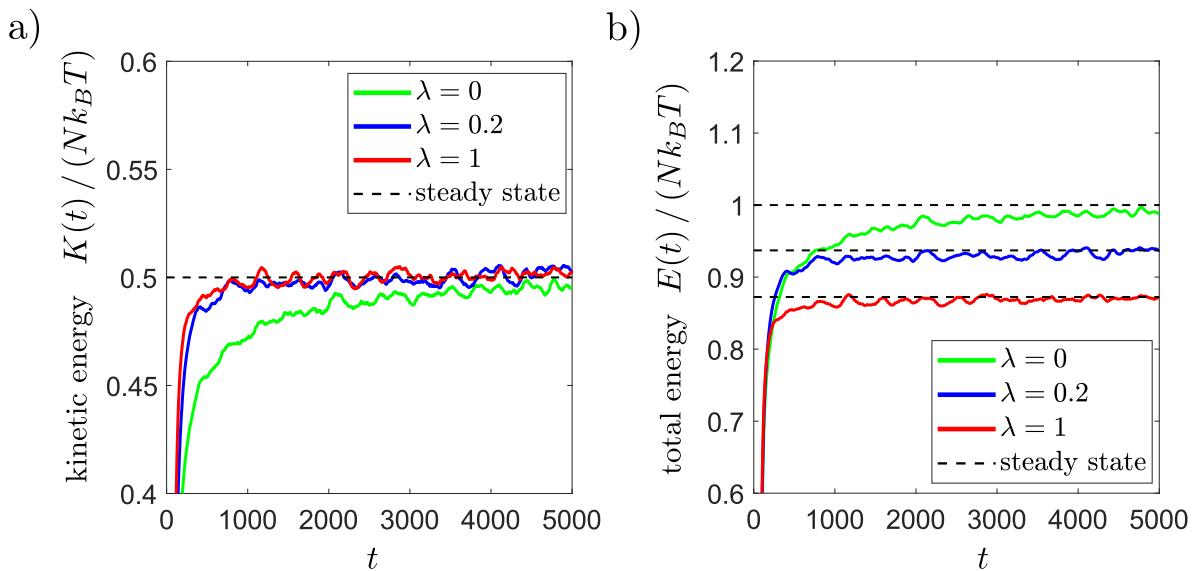}
\caption{Equilibration of total kinetic energy $K(t)=\sum_j p_j^2/2$ and total energy $E(t)$ ($N=32,\, k_BT=0.1$).  The straight (colored) lines are simulation for different strengths of the coupling while the black dotted lines display the value of the energy predicted from the partition function. The kinetic energy always reaches $Nk_B T/2$ independent of coupling, the total energy approaches $U(\lambda)$ and decreases with coupling. Nonlinearity accelerates the equilibration process.}
\label{fig:energy_time_evolution}
\end{figure}

\begin{figure}[t!]
\centering
\includegraphics[width=1\textwidth]{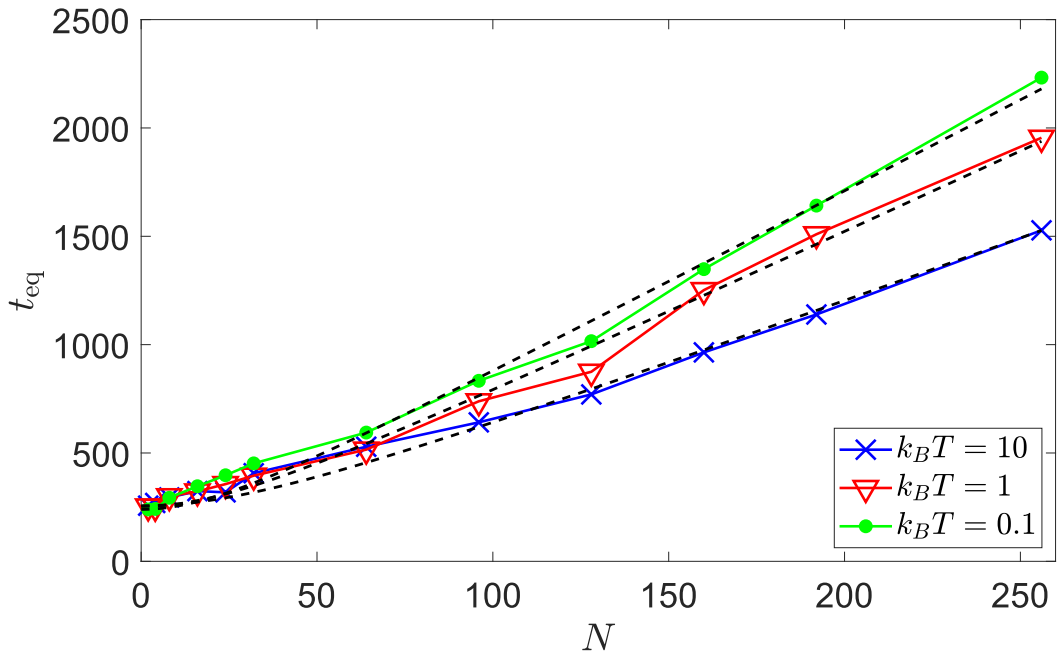}
\caption{Dependence of the equilibration time on system size. The equilibration time is linear in the system size $t_{\mathrm{eq}} \simeq \sqrt{t_0^2+t_1^2 N}$ for $N$ large enough (fixed $\lambda=10$). The single-body relaxation time is $t_0\approx 300$ in the plot. The slope depends on temperature $t_1=t_1(T)$ and increases with decreasing $T$, explicitly $t_1(k_B T=10)=5.8$, $t_1(k_B T=1)=7.5$ and $t_1(k_B T=0.1)=8.4$.}
\label{fig:equilibration time on system size}
\end{figure}

Having assessed the equilibrium properties of the system, next we utilize the stochastic 
model to numerically investigate the dependence of the relaxation to steady-state  
on the system size, temperature and nonlinearity. 
We define the equilibration time $t_{\mathrm{eq}}$ as the minimum time for which 
the total energy $E(t)$ reaches the equilibrium energy $U=U(\lambda)$ 
(\ref{eq:internal energy beta-LFPU}) and stays around it within fluctuations
of the order of a small fraction of $\sim k_BT$. 
Numerically, it is convenient to take a window $|E-U|/(k_BTN)<1\%$ 
(illustrated in figure \ref{fig:energy_time_evolution} b)).
\\
For a given temperature $T$ and coupling $\lambda$, we find that the equilibration 
time depends linearly on the system size, provided $N$ is large enough, 
see figure \ref{fig:equilibration time on system size}. 
We have fixed $\lambda=10$ (strong coupling) and plotted $t_{\mathrm{eq}}$ 
as a function of $N$ for various temperatures, finding good agreement with the 
following fit (the numerical values of the parameters are given in Appendix \ref{sec:appendix})

\begin{equation}
\label{eq:equilibration time dependent on system size}
t_{\mathrm{eq}}(N)\big|_{\lambda,T} \simeq \sqrt{t_0^2+t_1^2 N^2},
\end{equation} 

\begin{figure}[t!]
\centering
\includegraphics[width=1\textwidth]{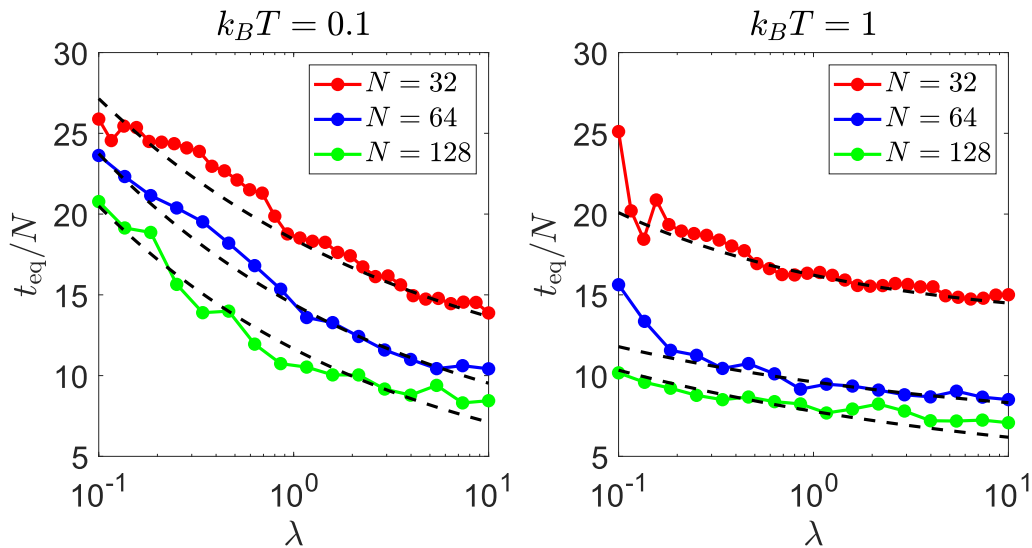}
\caption{Dependence of the equilibration time on coupling. The equilibration has a power-law dependence on the coupling $\sim\lambda^{-\mu}$ with exponent $\mu$ in the range of $[0.23-0.35]$ for the shown graphs.}
\label{fig:equilibration time on coupling }
\end{figure}

\noindent 
Besides being directly suggested by visual inspection of the numerical data, this
fit also responds to a simple physical interpretation.
Indeed, $t_0$ is associated to single-body relaxation time, while 
$t_1$ is the increment of $t_{\mathrm{eq}}$ due to the insertion of a single 
extra-node in the lattice chain. For large $N>100$, a linear scaling is clearly observed, with
coefficient $t_1=t_1(T)$ depending on the temperature, but not on the coupling strength. 
The main outcome of this analysis is the linear dependence of $t_\mathrm{eq}$ on $N$. 
\\
Figure \ref{fig:equilibration time on coupling } reports 
the equilibration time $t_{\mathrm{eq}}/N$ as a function of the nonlinearity $\lambda$. 
For a given temperature, the curves for different $N$ are equidistantly spaced, 
e.g. $t_{\mathrm{eq}}(\lambda,N=32)/32-t_{\mathrm{eq}}(\lambda,N=64)/64=\mathrm{const.}$ 
indicating that $t_1$ depends only weakly on the nonlinear coupling. 
For the case of a few sites $N\rightarrow 1$, temperature (at fixed $\lambda$) 
has no effect on the equilibration time, as seen from the fact that all curves 
in the figure \ref{fig:equilibration time on system size} merge into a single one 
for $N\rightarrow 1$).

\begin{figure}[b!]
\centering
\includegraphics[width=1\textwidth]{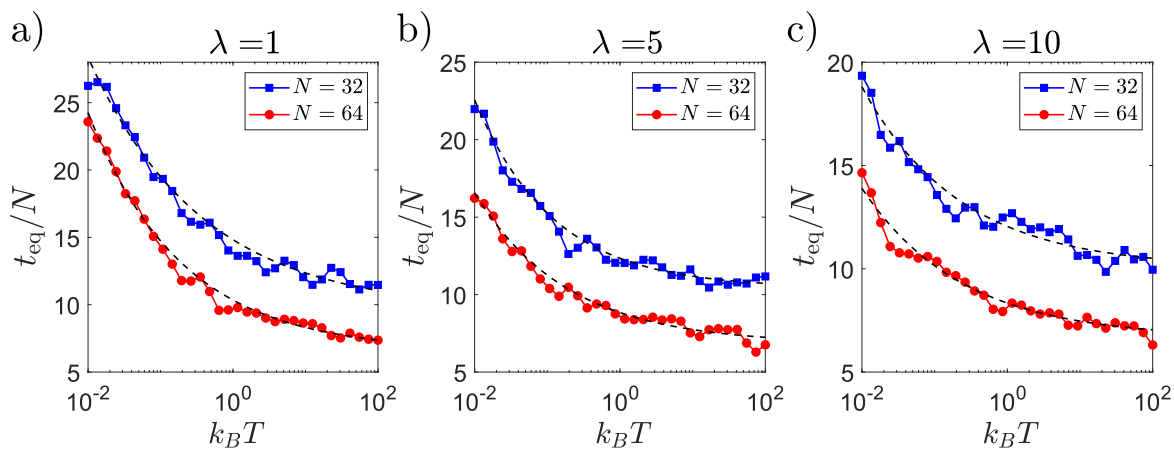}
\caption{Dependence of the equilibration time on temperature. We find a power-law behavior $\sim T^{-1/3}$ for various coupling strengths and system sizes.}
\label{fig:equilibration time as function of temperature}
\end{figure}

The linear dependence of $t_\mathrm{eq}$ on the system size is the main result of the work. 
We have added a cubic nonlinearity to the quartic FPUT model (\ref{eq:beta-LFPU}) and 
performed simulations to study $t_\mathrm{eq}(N)$ for fixed $\lambda$ and $T$ (not shown in this paper). 
The linear relationship is still valid in this case, as long as  the cubic nonlinearity 
is sufficiently weak to act as perturbation to the quartic model, namely as long as
the potential displays a single minimum. 

\begin{figure}[t!]
\centering
\includegraphics[width=1\textwidth]{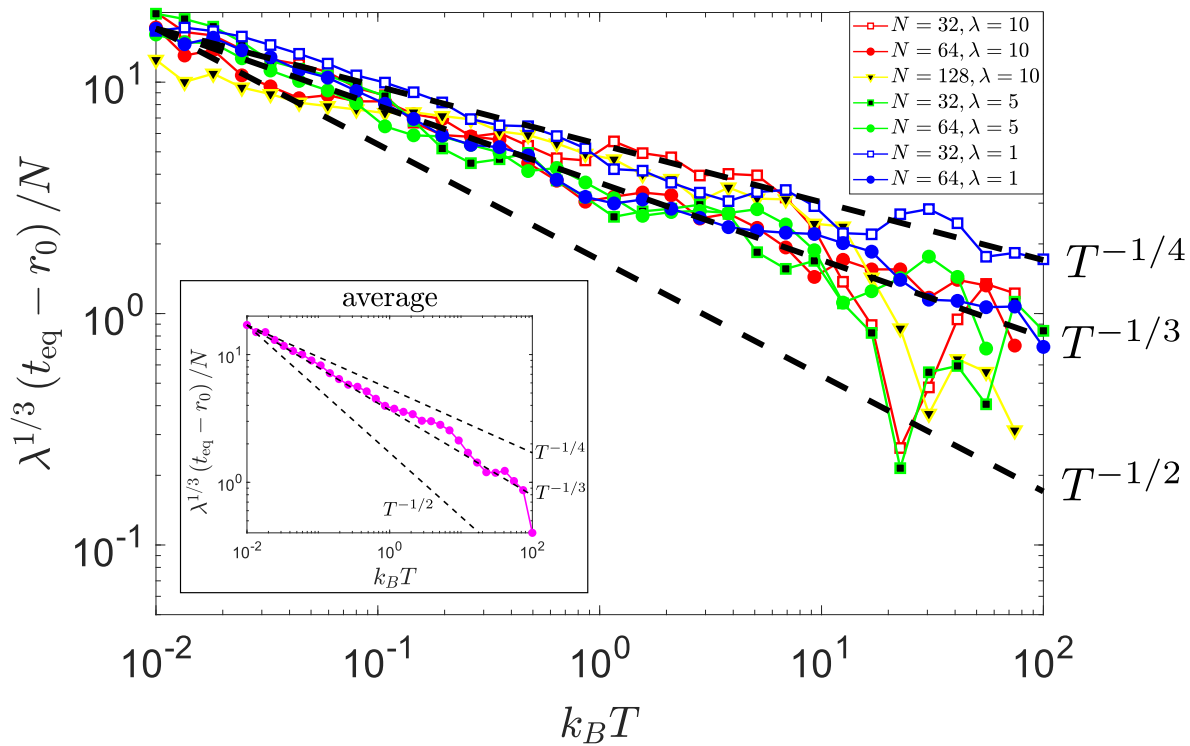}
\caption{Equilibration time as a function of temperature for different couplings and system sizes. The equilibration time follows a power-law with exponent $T^{-1/3}$ over many orders of magnitude of the thermal energy. In the left bottom panel we show the average of all curves, assuming they all follow the suggested power-law.}
\label{fig:master plot t_eq}
\end{figure}

Next, we analyse the relaxation time as a function of temperature 
(see figure \ref{fig:equilibration time as function of temperature}). 
The equilibration time decays weakly over a large range of simulated bath 
temperatures, and appears to be satisfactorly fitted by a power-law decay of the form: 

\begin{equation}
\label{eq:equilibration time dependent on temperature}
t_{\mathrm{eq}}(T)\big|_{N,\lambda} \simeq r_1\,{T}^{-1/3}+r_0.
\end{equation}  

\noindent Fitting more generally $t_{\mathrm{eq}} \simeq 
r_1\,{T}^{-\nu}+r_0$, the exponent $\nu\approx 1/3$ is 
found for different couplings and system sizes (see 
\ref{sec:appendix}). The first parameter $r_1$ decays 
weakly with $\lambda$ and increases linearly with $N$, as 
the two curves  for $N=32$ and $N=64$ in each subfigure 
\ref{fig:equilibration time as function of temperature} a) 
- c) are almost equidistantly spaced. The second parameter 
$r_0$ is a sublinear function of $N$ and numerically 
almost independent on $\lambda$, since the curves in figure 
\ref{fig:equilibration time as function of temperature} a) 
- c) (different $\lambda$, fixed $N$) have the same value 
for the largest bath temperature. We relate the parameters 
in (\ref{eq:equilibration time dependent on system size}) 
and (\ref{eq:equilibration time dependent on temperature}) 
in the large $N$ limit: the parameter $r_0$ takes 
numerically values similar to the case of the single-body 
relaxation time $t_0$ in our simulations at large $N$.
By expanding 
(\ref{eq:equilibration time dependent on system size}) in 
$N$, we identify the term $t_1 N$ with $r_1 T^{-1/3}$, thereby 
deducing $t_1\sim {T}^{-1/3}$. 
\\
At last, we investigate the effect of the coupling $\lambda$ on 
the equilibration time. From the spectral time evolution 
(see figure \ref{fig:distribution_of_modes}), we have already observed that 
equilibration is accelerated by nonlinearity. 
As in the temperature-dependent study, $t_{\mathrm{eq}}$ depends weakly on $\lambda$, 
a power-law with a small exponent

\begin{equation}
\label{eq: t_eq on lambda}
t_{\mathrm{eq}}(\lambda)\big|_{N,T}  \simeq u_1 \lambda^{-1/3}+u_0.
\end{equation} 

\noindent We have again assumed first a general dependence like 
$t_{\mathrm{eq}}(\lambda)  \simeq u_1 \lambda^{-\mu}+u_0$, and concluded 
from our data that $\mu\approx 1/3$ (see \ref{sec:appendix}). 
The time-asymptotic value $u_0$ does not significantly depend on $N$, and increases with 
$T$, assuming numerically comparable values like $r_0$ and $t_0$. 
Again, we can indentify them in the large $N$-limit and relate $u_0$ to $t_0$ 
and $r_0$, to give the leading contribution to the equilibration time in temperature 
(there might be also a contribution from $u_0$). 
It follows that $t_0$ and $r_0$ scale like $\sim \lambda^{-1/3}$. 
\\
The power-laws (\ref{eq:equilibration time dependent on temperature}) 
and (\ref{eq: t_eq on lambda}) can be motivated by dimensional analysis. The quantity

\begin{equation}
\omega_\mathrm{nl} = \left(\lambda k_B T\right)^{1/4}\,m^{-1/2} 
\end{equation} 

\noindent has the dimension of a inverse time. This suggests that if the equilibration time depends like a power-law on $\lambda$, 
then it should also depend like a power-law on $T$ with the same exponent, and vice versa. 
Like in thermal equilibrium, $\lambda$ plays the role of an inverse temperature. 
The other time-scales in the problem are the inverse dissipation constant, 
multiplied by the mass $\tau=\gamma/m$ and the harmonic frequency $\omega_\mathrm{har}=\sqrt{k/m}$. The simulations provide strong evidence for a behavior of the form
$t_\mathrm{eq}\sim(\lambda T)^{-1/3}$, see figure \ref{fig:master plot t_eq}, hence 
the nonlinear frequency enters like $t_\mathrm{eq}\sim \omega_\mathrm{nl}^{-4/3}$. 

\section{Discussion and conclusion}

We have implemented the quartic FPUT model and succesfully reproduced
the equilibrium canonical ensemble using Langevin baths. 
The equilibrium energy of our numerical approach agrees with the internal 
energy expected from the canonical ensemble (cf.~figure \ref{fig:energy}).
By an exact integration of the partition function of the nonlinear chain, we have been able to
recover non-perturbative results from statistical mechanics, covering 
both weak and strong coupling regimes. 
Numerical integration of the stochastic differential equations matched the expression of the
internal, nonlinear energy and harmonic energy from statistical mechanics in the time-asymptotic limit.
\\
It was found that the mentioned components of the equilibrium energy of the quartic FPUT, normalized 
to the thermal energy in a linear chain, depend only on the product of 
temperature and coupling via the dimensionless scaling $z=8k_B T\lambda$, 
i.~e.~$\left[U/(Nk_BT)\right](z)$.  
By the found scaling in $z$, we can make a proportionality argument: consider 
the potential of a single bond $V(\phi) = \frac{1}{2}\phi^2+\frac{1}{4}\lambda\phi^4$ of 
the quartic FPUT and a change in coupling $\lambda\rightarrow 2\lambda$. 
The internal energy of the bond remains exactly the same if we cool down the 
bond by $T\rightarrow T/2$, regardless of the interaction strength, temperature 
of the baths and also system size. At equilibrium, all bonds feel the bath as 
if it was adjacent to them. 
The amount of nonlinearity $\eta$ is likewise only a function of $z$, providing a 
closed formula (\ref{eq:nonlinear coupling}) to distinguish quantitatively 
between the strong and weak coupling regimes.
\\ 
Attaching the nonlinear system in the framework of Langevin baths puts us 
in the position to investigate the time evolution of the FPUT during the 
thermalization process. 
It is found that the nonlinearity accelerates equipartition, although not to a dramatic extent.
This becomes clear from comparing the classic FPUT with the harmonic chain 
in terms of energy distribution.
 In the harmonic chain, energy is kept only in the intially excited mode, whereas 
in the FPUT case, energy is distributed among all modes  
(provided at least one even and odd modes are initially excited in the quartic FPUT). 
 When thermal baths are attached to the nonlinear FPUT, energy equipartition 
is reached through the Langevin terms under the presence of nonlinearity, while in the 
linear chain, this is obtained  solely by the action of the Langevin terms.
\\
Nonlinear interaction adds to energy distribution 
among modes, thus speeding up the thermalization process.
 This happens selectively in different regions of the spectrum. 
Energy is faster channeled to the high modes as the coupling is amplified. 
The equipartition time is a function of system size, temperature and coupling strength. 
For $N$ large enough ($N>100$), it increases linearly with the system size, with 
the slope dependent on temperature but not on the coupling strength. 
The relaxation time is increased by increasing the temperature of the baths 
and displays a power law behavior with exponent $-1/3$. 
Increasing the coupling $\lambda$ leads to quicker equilibration, following 
again power-law dependence, with approximately the same exponent.
\\
In a future work, the faster distribution of energy, favoring high-oscillating modes, should 
be investigated by considering the transient solution of the mode equations. 
It would also be interesting to investigate the effect of dimensionality on the scaling of 
the equilibration time with system size. 
We hope that the stochastic implementation of the canonical ensemble presented in this paper
can prove useful to study equilibrium and non-equilibrium properties of 
similar nonlinear models, such as the Toda chain.
   
\section*{Acknowledgments}

This work is part of MUR-PRIN2017 project ”Coarse-grained description for non- equilibrium systems and transport phenomena (CO-NEST)” No. 201798CZL whose partial financial support is acknowledged. One of the authors (SS) acknowledges funding from the European Research Council under the Horizon 2020 Programme Grant Agreement n. 739964 ("COPMAT"). HS was financially supported by the ERASMUS program and the Physics Advanced program of the Elite Network of Bavaria (University of Regensburg, Germany), and is grateful for the hospitality of the Scuola Normale Superiore.

\newpage
\appendix
\section{Fitting parameters}
\label{sec:appendix}

We state fitting parameters used in figures \ref{fig:equilibration time on system size}, \ref{fig:equilibration time on coupling } and \ref{fig:equilibration time as function of temperature}.

\begin{table}[h!]
\centering
\begin{tabular}{c|c|c|c}
$k_B T$ & 0.1 & 1 & 10
\\
\hline
$t_1$ & 8.4 & 7.5 & 5.8
\end{tabular}
\caption{Fitting parameters for the dependence of the equilibration time on system size $t_{\mathrm{eq}}(N)=\sqrt{t_0^2+t_1^2 N^2}$ and $\lambda=10$. The single-body relaxation time is $t_0\approx 300$.}
\end{table}

\begin{table}[h!]
\centering
\begin{tabular}{c|p{4cm}|p{4cm}|p{3.5cm}}
 & $\lambda=1$ &  $\lambda=5$ & $\lambda=10$
\\
\hline
$N=32$ 
& $r_1=168,r_0=306$ $\nu=0.28$
&  $r_1=61.6,r_0=333$  $\nu=0.39$
& $r_1=64,r_0=321$ $\nu=0.32$
\\
\hline
$N=64$ 
& $r_1=242,r_0=420$ $\nu=0.34$
& $r_1=129,r_0=437$ $\nu=0.34$
&$r_1=107,r_0=426$ $\nu=0.32$
\end{tabular}
\caption{Fitting parameters for the dependence of the equilibration time on temperature $t_{\mathrm{eq}}(T)=r_1 T^{-\nu}+r_0$. The exponent $\nu$ is nearly independent of $\lambda$ and $N$, and close to $\frac{1}{3}$.}
\end{table}

\begin{table}[h!]
\centering00
\begin{tabular}{c|p{4cm}|p{4cm}|p{4cm}}
 & $N=32$ & $N=64$ & $N=128$
\\
\hline
$k_B T=0.1$ & $u_1=340, u_0=250$ $\mu=0.26$ & $u_1=659,u_0=264$ $\mu=0.28$ & $u_1=1192,u_0=300$, $\mu=0.29$
\\
\hline
$k_B T=1$ & $u_1=100,u_0=419$, $\mu=0.35$ &  $u_1=200,u_0=415$, $\mu=0.23$ & $u_1=826,u_0=206$, $\mu=0.34$
\end{tabular}
\caption{Fitting parameters for the dependence of the equilibration time on coupling $t_{\mathrm{eq}}(\lambda)=u_1 \lambda^{-\mu}+u_0$.}
\end{table}

\newpage
\section*{References}
\bibliography{bibliography/bib_lfpu}
\bibliographystyle{iopart-num}

\end{document}